\documentclass[10pt,conference]{IEEEtran} 
\IEEEoverridecommandlockouts

\usepackage{cite}
\usepackage{amsmath,amssymb,amsfonts}
\usepackage{algorithmic}
\usepackage{graphicx}
\usepackage{textcomp}
\usepackage{xcolor}
\usepackage{multirow}
\usepackage{booktabs}
\usepackage{makecell}

\usepackage[utf8]{inputenc}

\usepackage{CJKutf8}

\def\BibTeX{{\rm B\kern-.05em{\sc i\kern-.025em b}\kern-.08em
    T\kern-.1667em\lower.7ex\hbox{E}\kern-.125emX}}

\begin{document}

\title{Bridging the Modality Gap:\\
Softly Discretizing Audio Representation for LLM-based Automatic Speech Recognition\\

\thanks{This work was supported in part by the National Science Foundation under Grant awards CCRI \#2016725, EAGER \#2140415, Dev.Sci. \#2341384, and in part by Univ. of Texas at Dallas from the Distinguished University Chair in Telecommunications Engineering held by J. Hansen.}
}


\author{
\IEEEauthorblockN{Mu Yang, Szu-Jui Chen\textsuperscript{\dag}, Jiamin Xie\textsuperscript{\dag}, John H. L. Hansen}
\IEEEauthorblockA{
\textit{Centor for Robust Speech Systems (CRSS), University of Texas at Dallas, USA} \\
\{mu.yang, szu-jui.chen, jiamin.xie, john.hansen\}@utdallas.edu
}
}


\maketitle

\begin{abstract}
One challenge of integrating speech input with large language models (LLMs) stems from the discrepancy between the continuous nature of audio data and the discrete token-based paradigm of LLMs. To mitigate this gap, we propose a method for integrating vector quantization (VQ) into LLM-based automatic speech recognition (ASR). Using the LLM embedding table as the VQ codebook, the VQ module aligns the continuous representations from the audio encoder with the discrete LLM inputs, enabling the LLM to operate on a discretized audio representation that better reflects the linguistic structure. We further create a “soft discretization” of the audio representation by updating the codebook and performing a weighted sum over the codebook embeddings. Empirical results demonstrate that our proposed method significantly improves upon the LLM-based ASR baseline, particularly in out-of-domain conditions. This work highlights the potential of soft discretization as a modality bridge in LLM-based ASR.
\end{abstract}


\renewcommand{\thefootnote}{\dag}
\footnotetext{Contributed equally and are co-second authors.}

\section{Introduction}

Large Language Models (LLMs) \cite{touvron2023llama,brown2020language,hui2024qwen2,vicuna2023} have significantly advanced the field of Natural Language Processing.
Recently, there has been growing interest in extending LLMs to speech-related tasks \cite{gong2023listen,tangsalmonn,Qwen2-Audio,fathullah-etal-2024-audiochatllama,wang2024diarizationlm}.

In this work, we focus on the task of Automatic Speech Recognition (ASR). LLM-based ASR systems typically adopt a pipeline consisting of an audio encoder, a connector, and an LLM decoder \cite{10447605,ma2024embarrassingly,verdini2024connect,10389705,10445874,peng2024voicetextblender}. The connector down-samples and projects audio representations into a form compatible with the LLM input space, and the LLM subsequently decodes transcript tokens. 
Despite its simple yet effective design, a recent work \cite{11010998} has shown that SLAM-ASR \cite{ma2024embarrassingly}, a representative model in this paradigm, suffers significant performance degradation under cross-domain conditions (i.e., mismatched training and test domains), compared to non-LLM-based ASR systems. This highlights a key limitation of current LLM-based ASR systems: generalization remains an open challenge.

One major challenge in integrating speech input with LLMs is the modality gap between continuous audio representations and the discrete token representations used by LLMs. We approach this problem from a novel perspective: discretizing the continuous audio features to better align with the token-based paradigm of LLMs. LLMs are fundamentally designed to operate in a discrete space, where each token represents a distinct, quantized unit of information. In contrast, audio encoders produce smooth, high-dimensional latent representations that do not naturally align with the discrete prediction framework required by LLMs. 
To this end, we propose using Vector Quantization (VQ) to discretize the audio representations, enabling alignment with the finite set of discrete token embeddings in the LLM. Our goal is to obtain representations that are both compatible with the LLM’s input space and informative enough to preserve acoustic distinctions critical for ASR. For example, tokens with similar pronunciations should map to nearby points in the embedding space, while those with distinct pronunciations should be well separated. We hypothesize that such discretization helps the LLM better capture acoustic characteristics in its embedding space, thereby improving ASR generalization.

\begin{figure*}
  \centering
  \includegraphics[width=0.8\linewidth]{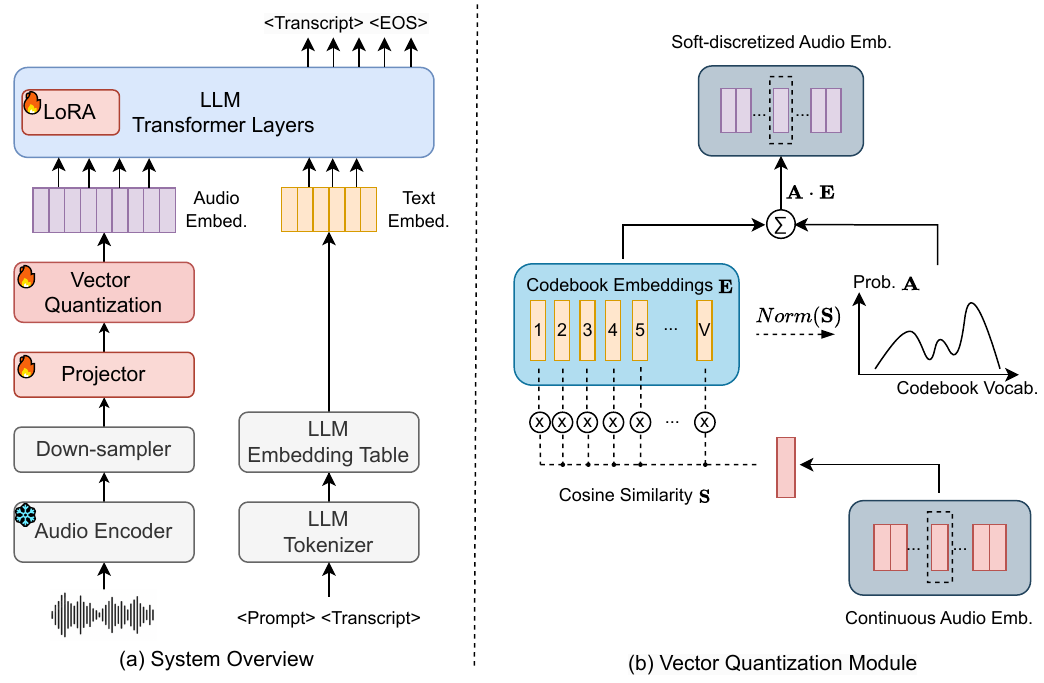}
  \vspace{-3mm}
  \caption{(a) Overall system architecture of the LLM-based ASR with a VQ module. (b) Design of the VQ module illustrating the soft discretization of audio representations. The module is trained in two stages: in the first stage, audio representations are aligned with LLM token embeddings using a fixed codebook initialized from the LLM embedding table; in the second stage, discretization is softened via a weighted sum, and the codebook is updated.}
  \label{fig:system}
\end{figure*}

We build upon the SLAM-ASR framework and introduce a VQ module between the projector and the LLM to produce discretized audio embeddings (Fig.~\ref{fig:system} (a)). Initial experiments reveal that when jointly trained with the rest of the model, the VQ module fails to learn a meaningful codebook that aligns with the LLM token embeddings. To address this issue, we adopt a two-stage training strategy. In the first stage, the VQ codebook is initialized with the LLM embedding table and kept fixed, while the rest of the model is updated to align audio representations with LLM embeddings. In the second stage, ``soft discretization'' is introduced via a weighted sum of codebook entries, and the codebook is updated to incorporate acoustic information. In both stages, the system is trained end-to-end using the same Cross Entropy loss as SLAM-ASR, without adding auxiliary loss terms. This training strategy allows the VQ module to learn a codebook that captures acoustic characteristics while remaining compatible with the LLM. Our empirical results show that the proposed method significantly improves upon the SLAM-ASR baseline, particularly in out-of-domain conditions, demonstrating improved generalization in LLM-based ASR systems. Additionally, extensive probing analyses confirm that the VQ module produces audio representations well-aligned with LLM text embeddings.

\section{Related Work}
Prior works on LLM-based ASR primarily focus on connector module architectures, aiming at reducing speech-text representation gap and mitigating sequence length discrepancy. Three connector desgins, including fully connected layers, multi-head cross-attention, and querying transformer were explored in \cite{10445874}.  Zhang et al.  \cite{zhang2025soundwave} proposed a three-stage training strategy incorporating alignment and length-shrinking adapters as connectors. Verdini et al. \cite{verdini2024connect} conducted a comparative study across different connector-LLM combinations. In contrast, we propose a discretization-based approach that aligns continuous audio representations with the discrete token paradigm of LLMs. Prior works exist using \textit{offline} K-means for quantization \cite{xu24d_interspeech, wang2024comparative}, but we employ an \emph{online} VQ module jointly trained with the LLM. Our method is related to ASQ \cite{10374223}, which uses Gumbel-Softmax VQ in a pre-quantized encoder-decoder ASR framework, but differs in targeting LLM-based ASR and supporting end-to-end training. Additionally, while Yang et al. \cite{yang2023generative} used LLMs for post-hoc correction of ASR N-best hypotheses, our approach directly predicts transcriptions from audio input.

\section{Method}

\subsection{System Overview}
Our proposed model, shown in Fig.~\ref{fig:system}, is based on the SLAM-ASR framework \cite{ma2024embarrassingly}, which consists of an Audio Encoder, a Down-sampler, a Projector, and an LLM decoder. 
The Audio Encoder takes raw audio or Mel spectrogram as input and outputs continuous representations. Following SLAM-ASR, we apply the same Down-sampler operation, which stacks 5 consecutive audio frames into a single frame, reducing the temporal resolution by a factor of 5. The Projector, composed of two linear layers with a ReLU activation, then maps the downsampled representations to audio embeddings with the same dimension as the LLM input embeddings. 

During training, the system input follows a prompt template: ``USER: [audio input] Transcribe speech to text. ASSISTANT: [transcript].'' The audio input is processed by the Audio Encoder, Down-sampler, and Projector to generate the audio embeddings, while the transcript and other template texts are tokenized and embedded using the LLM embedding table to form the text embeddings. 
The LLM then takes both embeddings as input. The entire system is trained end-to-end using Cross Entropy loss and causal masking to predict the target transcript tokens. During inference, no ground-truth transcript is provided; transcript tokens are generated autoregressively. Beam Search with a beam size of 4 is used for decoding.

As shown in Fig.~\ref{fig:system}, our proposed VQ module is inserted between the Projector and the LLM Transformer layers to discretize the audio embeddings before they are fed into the LLM. We also apply a low-rank adaptation (LoRA) \cite{hu2022lora} module to the LLM to enable efficient fine-tuning.

\subsection{Modality Alignment via Vector Quantization}
We propose to use Vector Quantization (VQ) to discretize the audio representations. The VQ module, together with the other modules, is trained by a 2-stage strategy. The first stage focuses on aligning the continuous audio representation with the LLM embedding table for the discrete text tokens, and the second stage creates ``softened'' audio representations using a weighted sum of the VQ codebook entries to account for acoustic information.

\subsubsection{\textbf{Stage 1: Aligning Audio Representation with LLM Token Embedding}} \label{sec:stage1}
In the first stage, the VQ module is trained with a \emph{fixed} codebook initialized with the LLM embedding table. For each audio embedding, the quantization operation is performed by selecting the codebook entry that has the highest cosine similarity. Formally, let \(\mathbf{z} \in \mathbb{R}^{T \times D}\) be the continuous audio embeddings output from the Projector, where \(T\) is the number of frames and \(D\) is the feature dimension. The codebook is defined as \(\mathbf{E} \in \mathbb{R}^{V \times D}\), where \(V\) is the number of codebook entries (i.e. LLM vocabulary size). 
The cosine similarity scores between each audio embedding \(\mathbf{z}_{t}\) at frame $t$ and each codebook entry \(\mathbf{e}_j\) are then computed.
The quantized audio embedding \(\mathbf{q}_{t}\) for \(\mathbf{z}_{t}\) is obtained by selecting the codebook entry \(\mathbf{e}_j\) that has the highest cosine similarity:
\begin{equation}
  \mathbf{q}_{t}=\mathbf{e}_{j}, \quad \text { where } j=\operatorname{argmax}_{v} \operatorname{sim}(\mathbf{z}_{t}, \mathbf{e}_v)
\end{equation}
where \(\operatorname{sim}(\cdot, \cdot)\) denotes the cosine similarity function.
Due to the non-differentiable nature of $\operatorname{argmax}$ operation, we use the Straight-Through Estimator (STE) \cite{van2017neural,bengio2013estimating} to backpropagate the gradients during training:

\begin{equation}
  \tilde{\mathbf{z}}_{t} = \mathbf{z}_{t} + \operatorname{sg} \left[ \mathbf{q}_{t} - \mathbf{z}_{t} \right]
  \label{eq:ste}
\end{equation}
where \(\operatorname{sg}[\cdot]\) denotes the stop gradient operation. In the forward pass, $\tilde{\mathbf{z}}_{t} = \mathbf{q}_{t}$ is used, while in the backpropagation the gradients w.r.t VQ output $\tilde{\mathbf{z}}_{t}$ are identically passed to the original continuous audio embedding $\mathbf{z}_{t}$, since $\operatorname{sg} \left[ \mathbf{q}_{t} - \mathbf{z}_{t} \right]$ is seen as a constant.

Note that the quantized audio embeddings $\tilde{\mathbf{z}}$ are essentially LLM embedding table entries, in order to correctly perform ASR, the LLM requires $\tilde{\mathbf{z}}$ to provide information about the target tokens, thus aligning the audio representation with the LLM text embeddings. After stage one training, we expect audio embeddings $\tilde{\mathbf{z}}$ to have high cosine similarities with the text embeddings corresponding to the target tokens in the ground-truth transcript, while having low cosine similarities with irrelevant text tokens. 

\subsubsection{\textbf{Stage 2: Softening Discretization and Updating Codebook}} \label{sec:stage2}
In the second stage, we aim to integrate acoustic information into $\tilde{\mathbf{z}}$ by softening the discretization of audio embeddings and making the codebook entries trainable. For this purpose, we adopt a similar yet simpler approach to Gumbel-Softmax VQ \cite{jang2017categorical}. As shown in Fig.~\ref{fig:system} (b), we create a soft discretization of the audio representation:

\begin{equation}  
  \tilde{\mathbf{z}}_{t}^{soft} = \sum_{j=1}^{V} A_{t,j} \cdot \mathbf{e}_j
  \label{eqn:k}
\end{equation}
where \(\mathbf{A}_{t} \in \mathbb{R}^{V} \) is a cosine similarity-based distribution over the codebook entries:
\begin{equation}
  A_{t,j} = \frac{\exp(\operatorname{sim}(\mathbf{z}_{t}, \mathbf{e}_j))}{\sum_{j'=1}^{V} \exp(\operatorname{sim}(\mathbf{z}_{t}, \mathbf{e}_{j'}))}
\label{eq:softmax}
\end{equation}
This allows us to compute a weighted sum of the codebook entries, where the weights are determined by the cosine similarity scores between the audio embeddings and the codebook entries. This process can also be viewed as Attention mechanism \cite{vaswani2017attention}, where the audio embeddings \(\mathbf{z}\) serve as queries, the codebook \(\mathbf{E}\) serves as keys and values, allowing gradients to flow through the codebook entries and making them trainable. Such soft discretization retains more acoustic information while still aligning with the LLM text embeddings. 
We further experiment with the number of updatable codebook entries \(k\). In practice, we keep the top-\(k\) entries in \(\mathbf{A}_{t}\) and set the rest to zero in Equation~\ref{eqn:k}, so that only top-\(k\) entries are considered and gradients w.r.t the non-top-\(k\) entries are zeroed out. 

In addition, we also investigate the effect of soft discretization compared to hard discretization. For the latter, we quantize \(\mathbf{A}_{t}\) to an one-hot vector \(\mathbf{A}_{t}^{(one-hot)}\) by \(\operatorname{argmax}\), i.e. \(A_{t,j}^{(one-hot)} = 1\) if \(j = \operatorname{argmax}_{i}A_{t,i}\) and \(0\) otherwise. The hard-quantized audio embedding is then computed as:
\begin{equation}
  \tilde{\mathbf{z}}_{t}^{hard} = \sum_{j=1}^{V} A_{t,j}^{(one-hot)} \cdot \mathbf{e}_j
\end{equation}
A STE (Equation~\ref{eq:ste}) is used to backpropagate gradients to \(\mathbf{A}_{t}\) (and codebook entries):
\begin{equation}
  \tilde{\mathbf{A}}_{t} = \mathbf{A}_{t} + \operatorname{sg} \left[ \mathbf{A}_{t} ^{(one-hot)} - \mathbf{A}_{t} \right]
  \label{eq:ste2}
\end{equation}
Similarly, for hard discretization, we can control the number of updatable codebook entries in each backpropagation step by keeping top-\(k\) entries in \(\mathbf{A}_{t}\).

\begin{table*}[t]
  \caption{WER(\%)$\downarrow$  on Librispeech test-clean and CommonVoice 17.0 test set (denoted as cv17). \(\infty\) denotes WER greater than 100. Hard/soft discretization refers to using a single codebook entry/using a weighted sum of codebook entries. \(\mathbf{k}\) denotes the number of codebook entries used in the weighted sum, which also determines the number of updatable entries during backpropagation. See Sec.~\ref{sec:stage2} for more details.}
  \centering
\resizebox{0.95\textwidth}{!}{
\begin{tabular}{ccccccccc}
\toprule
\multirow{2}{*}{\textbf{ID}} & \multirow{2}{*}{\textbf{Audio Encoder}} & \multirow{2}{*}{\textbf{Model}} & \multirow{2}{*}{\textbf{LLM setting}} & \multicolumn{3}{c}{\textbf{Discretization setting}} & \multirow{2}{*}{\textbf{test-clean}} & \multirow{2}{*}{\textbf{cv17}} \\
\cmidrule(lr){5-7}
& & & & \textbf{Hard/Soft?} & \textbf{Codebook} & \(\mathbf{k}\) & & \\
\midrule
1 & \multirow{10}{*}{WavLM Large} & \multirow{2}{*}{SLAM-ASR\cite{ma2024embarrassingly}} & frozen & N/A & N/A & N/A & \textbf{3.73} & 68.27 \\
2 & & & LoRA r=32 & N/A & N/A & N/A & 3.78 & 46.12 \\
\cmidrule(lr){3-9}
3 & & Stage 1 & LoRA r=32 & hard & frozen & N/A & 7.88 & 43.40 \\
\cmidrule(lr){3-9}
4 & & \multirow{7}{*}{\quad+ Stage 2} & \multirow{7}{*}{LoRA r=32} & \multirow{4}{*}{hard} & \multirow{4}{*}{trainable} & All & 8.18 & 40.64 \\
5 & & & & & & 100 & 5.33 & 33.47 \\
6 & & & & & & 10 & 4.95 & 32.32 \\
7 & & & & & & 1 & 7.16 & 36.43 \\
\cmidrule(lr){5-9}
8 & & & & \multirow{3}{*}{soft} & \multirow{3}{*}{trainable} & All & \(\infty\) & \(\infty\) \\
9 & & & & & & 100 & \textbf{3.74} & \textbf{29.04} \\
10 & & & & & & 10 & 3.91 & 30.78 \\
\midrule
11 & \multirow{4}{*}{Whisper Medium} & \multirow{2}{*}{SLAM-ASR\cite{ma2024embarrassingly}} & frozen & N/A & N/A & N/A & 6.06 & 35.89 \\
12 & & & LoRA r=32 & N/A & N/A & N/A & 7.54 & 29.86 \\
\cmidrule(lr){3-9}
13 & & Stage 1 & LoRA r=32 & hard & frozen & N/A & 10.07 & 34.96 \\
\cmidrule(lr){3-9}
14 & & \quad+ Stage 2 & LoRA r=32 & soft & trainable & 10 & \textbf{5.63} & \textbf{25.53} \\
\bottomrule
\end{tabular}
}
\label{tab:main}
\vspace{-3mm}
\end{table*}

\subsection{Training}
In stage 1, only the Projector and the LLM LoRA parameters are trained. In stage 2, the VQ codebook, as well as the Projector and LLM LoRA parameters are trained together. We use the trained model from stage 1 to initialize stage 2. We trained multiple stage 2 models with different VQ settings (soft vs. hard discretization, with different \(k\) values), as discussed in Sec.~\ref{sec:stage2}.
In both stages the entire model is trained end-to-end with a Cross Entropy loss. We did not add commitment loss and codebook loss as in \cite{baevskivq,van2017neural}, since we found that the model works fine without these additional losses.

\section{Experimental Setup}

\textbf{Dataset}. We use Librispeech \cite{panayotov2015librispeech} to train all models.
Specifically, we train on the 960-hour Librispeech training set and evaluate word error rate (WER) on the test-clean set. For cross-domain evaluation, we use the English split of the CommonVoice 17.0 test set \cite{ardila2019common} (cv17). Since CommonVoice transcripts are often unnormalized and include non-standard punctuation and symbols, we apply Whisper's \texttt{BasicTextNormalizer} to both reference and hypothesis texts before computing WER. Additionally, we use a custom pre-cleaning function to preserve content inside parentheses (e.g., ``(laughter)''), which the Whisper normalizer would otherwise remove, and to replace ampersands with the word ``and'' for semantic consistency. This ensures a more accurate and fairer evaluation of ASR performance.

\textbf{Implementation Details}
Our model is implemented based on SLAM-ASR \cite{ma2024embarrassingly}. The VQ codebook is initialized with the LLM embedding table and made trainable during Stage 2 (Sec.~\ref{sec:stage2}). In Stage 1, we train for 10 epochs using the Adam optimizer with a peak learning rate of 1e-4. The warmup is set to 1,000 steps, after which the learning rate is held constant. Gradient accumulation is used to achieve an effective batch size of 32. In Stage 2, we use the same learning rate schedule but reduce the peak learning rate to 1e-5 and train for 2 epochs with an effective batch size of 16. We use Qwen2.5-0.5b \cite{hui2024qwen2} as the LLM, which has a vocabulary size of 151,936. A low-rank adaptation (LoRA) module with rank 32 is applied to the LLM to enable efficient fine-tuning.\footnote{Our initial experiments show that without LLM LoRA, the proposed model struggles to improve during training.} For the Audio Encoder, we experiment with two variants: WavLM Large \cite{9814838}, pre-trained using self-supervised learning (SSL), and Whisper Medium (\texttt{medium.en}) \cite{radford2023robust}, trained with supervised ASR objectives.

\section{Results}

We compare the proposed models with the SLAM-ASR baseline. Since our method fine-tunes the LLM using LoRA, while the original SLAM-ASR keeps the LLM frozen, we apply LoRA to the LLM in our SLAM-ASR baselines for a fairer comparison. Results are shown in Table~\ref{tab:main}. Comparing Row 1 vs. Row 2 and Row 11 vs. Row 12, we can see that enabling LoRA fine-tuning improves cross-domain WER on cv17, while maintaining similar (WavLM Large) or slightly worse (Whisper Medium) performance on the in-domain test-clean set. However, a substantial performance gap between in-domain and cross-domain settings remains, particularly for the WavLM Large encoder. 

\subsection{The Effect of different discretization strategies}
The results of our proposed method using soft discretization and an updatable codebook is presented in Rows 8-10 and Row 14 of Table~\ref{tab:main}. Comparing Rows 8-10 with Row 2, we observe that our method significantly outperforms the SLAM-ASR baseline on cv17, while achieving comparable performance on test-clean. A similar trend appears in Rows 14 and 12. This confirms that our method substantially improves the cross-domain generalization ability of LLM-based ASR systems.

Additionally, for both WavLM Large and Whisper Medium, the Stage 2 models outperform their corresponding Stage 1 model (Row 3 and Row 13), suggesting that ``softening'' the discretization is effective for both in-domain and cross-domain performance. Comparing the best-performing models for WavLM Large (Row 9) and Whisper Medium (Row 14), we find that the WavLM Large-based model performs better on test-clean, while the Whisper Medium-based model performs better on cv17. 

We further compare hard and soft discretization strategies in Stage 2 (Rows 4-7 and Rows 8-10), using the same stage 1 model as initialization (Row 3). The soft discretization strategy consistently outperforms hard discretization, suggesting it yields more informative and LLM-compatible audio representations, leading to improved ASR accuracy. However, when \(\mathbf{k}=\text{All}\), i.e., when the weighted sum includes all codebook entries and the entire codebook is updated during backpropagation, soft discretization degrades sharply. This is likely due to the large size of the LLM embedding table ($\sim$ 150k entries), which causes the weighted sum to become noisy and less informative, resulting in noisy gradients that disrupt codebook learning. 

Finally, comparing Rows 4-7 with Row 3, we observe the benefits of using a trainable codebook. Regarding the number of updatable codebook entries \(\mathbf{k}\), there appears to be a sweet spot for both hard and soft discretization: increasing or decreasing \(\mathbf{k}\) does not lead to consistent improvement or degradation. In general, a moderate number of trainable codebook entries (10-100) is sufficient to soften discretization, preserve acoustic information, and maintain alignment with LLM text embeddings.

\subsection{Embedding visualization}
\begin{figure}
    \centering
    \includegraphics[width=0.82\linewidth]{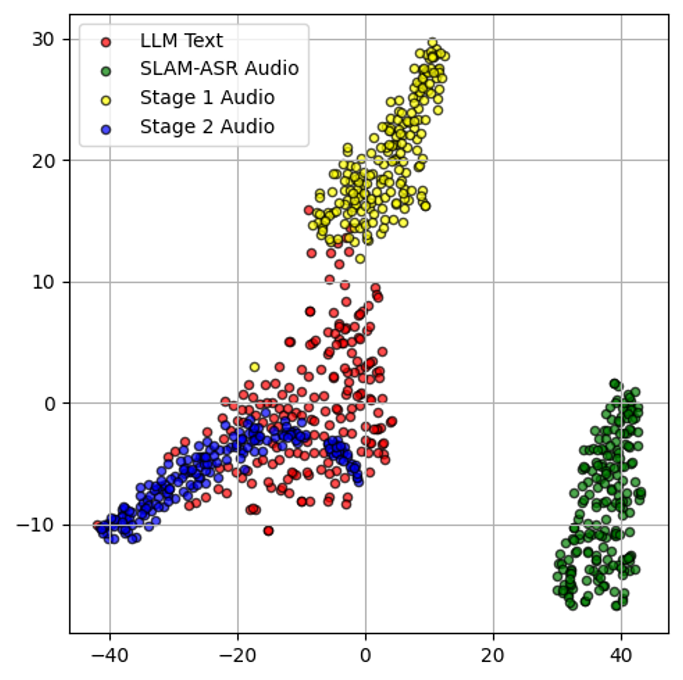}
    \caption{Embedding visualization. Red dots are text embeddings from the LLM embedding table. Green dots are audio embedding outputs of the Projector from the SLAM-ASR baseline (i.e. Row 2 in Table~\ref{tab:main}). Yellow and blue dots are audio embedding outputs of the VQ module from our proposed stage 1 and stage 2 model, respectively (Rows 3 and 9 in Table~\ref{tab:main}).}
    \label{fig:tsne}
    \vspace{-0.3cm}
\end{figure}

To demonstrate that the proposed method effectively aligns the audio embeddings with the LLM text embeddings, we visualize the audio and text embeddings in a 2D space using t-SNE \cite{van2008visualizing}. We randomly sample 200 utterances from the CommonVoice 17.0 test set and extract audio embeddings (i.e., the output of the VQ module) from the WavLM Large-based stage 2 and stage 1 models (Rows 9 and 3 in Table~\ref{tab:main}). For the baseline audio embeddings, we use the output of the Projector from the SLAM-ASR model (Row 2). Ground-truth transcripts are embedded using the LLM embedding table to obtain text embeddings. To obtain utterance-level representations, we average both audio and text embeddings along the time dimension. Fig.~\ref{fig:tsne} shows the 2D t-SNE visualization of these utterance-level embeddings.

We observe that the audio embeddings from the SLAM-ASR baseline (green dots) form a cluster that is far from the text embeddings (red dots). In contrast, the audio embeddings from our stage 1 model (yellow dots) are notably closer to the text embeddings, indicating that the VQ module effectively bridges the modality gap between audio and text. Moreover, after applying soft discretization and updating the codebook in stage 2, the audio embeddings (blue dots) show even greater overlap with the text embeddings, suggesting improved alignment between modalities.

\begin{figure}
  \centering
  \includegraphics[width=\linewidth]{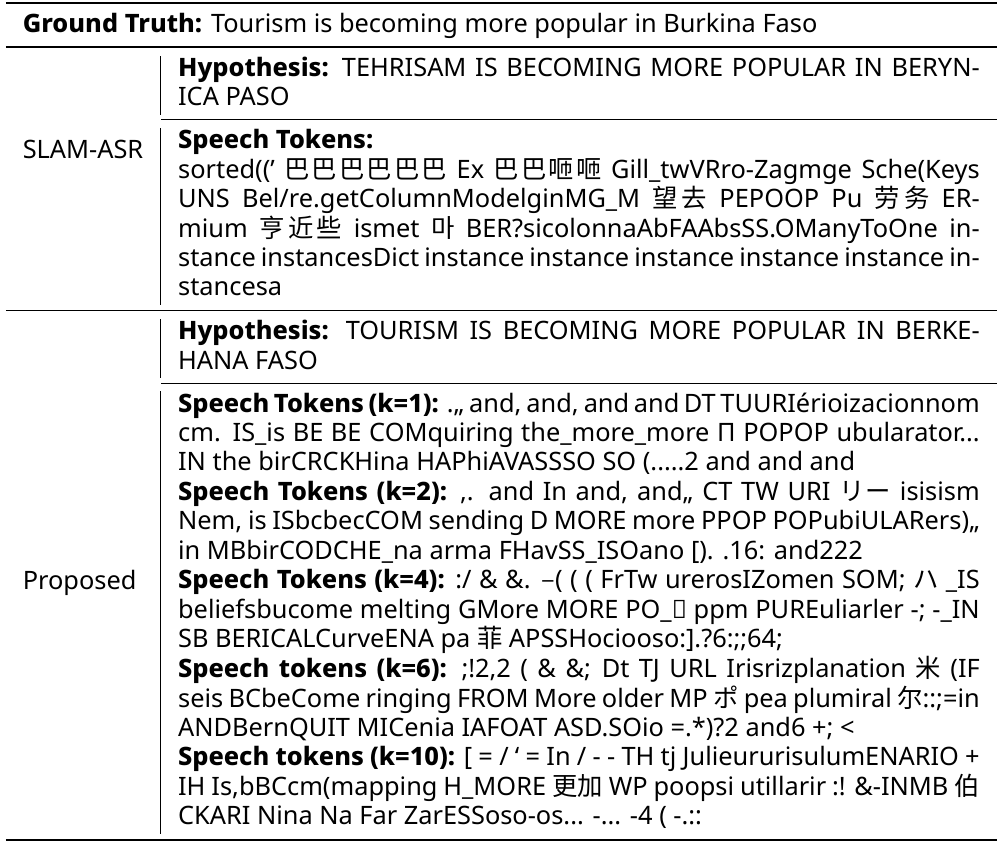}
  \caption{Ground truth, hypothesis, and the text tokens that have the highest cosine similarities with the audio embeddings. For the proposed model (Row 9 in Table~\ref{tab:main}), we show the top-k text tokens. For the SLAM-ASR baseline (Row 2 in Table~\ref{tab:main}), we show the top-1 text token.}
  \label{fig:sentence_example}
  \vspace{-0.3cm}
\end{figure}

\begin{table*}[t]
  \caption{Comparison of Hard or Soft discretization in stage 1. We show the WER (\%) on LibriSpeech test-clean and CommonVoice 17.0 test set (cv17) for different encoder and training settings. We use LLM with LoRA r=32 and k=10 for all soft discretization in this table.}
  \centering
  \resizebox{0.7\textwidth}{!}{
  \begin{tabular}{cccccc}
  \toprule
  \textbf{ID} & \textbf{Audio Encoder} & \textbf{Model Setting} & \textbf{Codebook} & \textbf{test-clean} & \textbf{cv17} \\
  \midrule
  1 & \multirow{5}{*}{WavLM Large} & Stage 1 - hard & frozen & 7.88 & 43.40 \\
  \cmidrule(lr){4-6} 
  2 & & \multirow{2}{*}{\quad + Stage 2 - soft} & frozen & 4.12 & 31.03 \\
  3 & & & trainable & 3.91 & 30.78 \\
  \cmidrule(lr){3-6}
  4 & & Stage 1 - soft & frozen & 5.43 & 43.75 \\
  5 & & \quad + Stage 2 - soft & trainable & 4.86 & 37.06 \\
  \midrule
  7 & \multirow{4}{*}{Whisper Medium} & Stage 1 - hard & frozen & 10.07 & 34.96 \\
  8 & & \quad + Stage 2 - soft & trainable & 5.63 & 25.53 \\
  \cmidrule(lr){3-6}
  9 & & Stage 1 - soft & frozen & 8.68 & 32.57 \\
  10 & & \quad + Stage 2 - soft & trainable & 7.28 & 30.88 \\
  \bottomrule
  \end{tabular}
  }
  \label{tab:codebook_ablation}
  \end{table*}

\subsection{What tokens are the audio embeddings aligned with?}
To further understand what the VQ module learned, we analyze the linguistic structure of the audio embeddings from the proposed model. Specifically, we inspect which text tokens are most similar to the audio embeddings. To do this, we sample an utterance from CommonVoice 17.0 test set and extract its audio embedding sequence. For each audio embedding in the sequence, we retrieve the top-\(\mathbf{k}\) text tokens from the LLM embedding table that have the highest cosine similarity with it. We refer to these as \textit{speech tokens}, as they are retrieved using speech representations as queries.

The speech tokens, along with the ground-truth transcript and model hypotheses, are shown in Fig.~\ref{fig:sentence_example}. We observe that the speech tokens retrieved by the proposed model (Row 9 in Table~\ref{tab:main}) are either semantically aligned with or acoustically similar in pronunciation to the text tokens in the ground-truth transcript. On the other hand, the SLAM-ASR baseline (Row 2 in Table~\ref{tab:main}) yields speech tokens that are mostly irrelevant. This indicates that our proposed method learns audio representations that better align with linguistic structure while also preserving important acoustic characteristics of the speech input.

Interestingly, although the training data (Librispeech) is purely English, the proposed model can retrieve non-English tokens that have similar pronunciations. 
\begin{CJK}{UTF8}{gbsn}
For example, the Chinese token \texttt{菲} has a pronunciation of ``fay'' and \texttt{尔} is pronounced ``er'', 
\end{CJK}
both of which correspond to syllables in the ground-truth transcript. This again demonstrates that the learned audio representations align well with the LLM embedding space across languages, highlighting the model’s ability to capture phonetic-level information.

\subsection{Discussion: The alternatives to hard-to-soft discretization}
In this work, we follow a two-stage training strategy: the first stage strictly discretizes the audio representations using the LLM embedding table, and the second stage softens the discretization and updates the codebook. We conduct an ablation study to investigate alternatives to this hard-to-soft scheme. 
Specifically, we train a stage 1 model using the soft discretization method described in Sec.~\ref{sec:stage2}, but keep the VQ codebook fixed. Then, we use this soft-discretization model to initialize and train a stage 2 model with the same settings: soft discretization and a trainable codebook. We set \(\mathbf{k} = 10\) for both stages. A comparison between this soft-to-soft alternative and the hard-to-soft scheme is shown in Table~\ref{tab:codebook_ablation}.

We observe that the hard-to-soft scheme (Rows 2-3 and 8) consistently outperforms the soft-to-soft alternatives (Rows 5 and 10) on both in-domain and cross-domain WER. Interestingly, while soft discretization in stage 1 yields better performance than hard discretization alone (Row 4 vs. Row 1 and Row 9 vs. Row 7), using this soft-discretized model to initialize stage 2 results in worse performance (Row 5 vs. Row 3 and Row 10 vs. Row 8). One possibility is that soft discretization in stage 1 does not enforce alignment with \emph{individual} LLM token embeddings, but instead aligns audio embeddings with a \emph{weighted sum} of token embeddings. This can lead to a less informative cosine-similarity distribution (\(\mathbf{A}_{t}\) in Equation~\ref{eq:softmax}), which may negatively affect codebook learning in stage 2. 
These findings suggest that hard discretization in stage 1 helps to enforce precise alignment between audio representations and individual LLM tokens, thereby enhancing the effectiveness of soft discretization in stage 2. 

We also consider another alternative: hard-to-hard discretization, as shown in Rows 3–7 of Table~\ref{tab:main}. The key difference from our proposed hard-to-soft discretization lies in how gradients are used to update the codebook in stage 2. The hard-to-hard method uses the STE (Equation~\ref{eq:ste2}) to backpropagate gradients from individual codebook entries, while the hard-to-soft method uses standard gradients from the weighted sum of entries. As shown in Table~\ref{tab:main}, our hard-to-soft strategy achieves greater improvements over the stage 1 model compared to hard-to-hard discretization. Disabling the codebook update in stage 2 of the hard-to-soft method (Row 2, Table~\ref{tab:codebook_ablation}) results in only a slight performance drop relative to the full version (Row 3), yet still significantly outperforms the hard-to-hard method (Row 1 in Table~\ref{tab:codebook_ablation}). This suggests that the performance gain of the hard-to-soft approach is mainly attributed to soft discretization, while codebook updating plays a secondary role. In contrast, the codebook update is essential in the hard-to-hard setup, as evidenced by Rows 3–7 in Table~\ref{tab:main}.

\section{Conclusion}
In conclusion, we presented a novel training method for LLM-based ASR, improving its generalization performance. We proposed using a vector quantization (VQ) module to mitigate the modality gap between the continuous audio representation space and the discrete LLM inputs. Our two-stage training method first strictly discretizes speech representations to the pre-trained LLM vocabulary embeddings and then enables a refinement of the quantization codebook. Experimental results demonstrate that our method of quantization significantly enhances the out-of-domain performance of LLM-based ASR systems without compromising in-domain performance. The probing studies showed a closer alignment between the quantized speech representations and the LLM text embedding space. Lastly, we decoded the speech tokens from the LLM inputs, which textually reveal the transition mechanism from acoustic to linguistic structures, highlighting the LLM's interpretation of speech in LLM-based ASR.

\newpage

\bibliographystyle{IEEEtran}
\bibliography{mybib}

\end{document}